\documentstyle[aps,multicol]{revtex}

\def\e{\varepsilon}
\renewcommand{\narrowtext}{\begin{multicols}{2} \global\columnwidth20.5pc}
\renewcommand{\widetext}{\end{multicols} \global\columnwidth42.5pc}
\begin{document}

\title{On nonuniversal conductance quantization in high-quality
quantum wires
}

\author{Anton Yu. Alekseev$^{* \dagger }$,
 Vadim V. Cheianov$^{* \dagger }$ }

\address{$^*$
Institute of Theoretical Physics, Uppsala University,
Box 803, S-75108, Uppsala, Sweden}

\address{$^\dagger $
Steklov Mathematical Institute, Fontanka 27, 191011,
St.Petersburg, Russia}

\date{December 1996}
\maketitle

{\tightenlines
\begin{abstract}
We present a theoretical analysis of recent experimental results of
Yacoby {\em et al.} on transport properties of high quality
quantum wires. We suggest an explanation of observed deviations
of the conductance from the universal value $2e^2/h$ per
channel in the wire. We argue that at low temperatures and
biases the deviation can be a consequence of anomalously
enhanced backscattering of electrons entering the 2DEG from
the wire and is not connected to intrinsic properties of 1DEG.

\vskip 0.1cm

\end{abstract}
}


\narrowtext

Recently Yacoby {\em et al.} \cite{Yacoby} presented measurements of
conductance and differential conductance of  high quality quantum wires
prepared using cleaved edge overgrowth technique. Despite the recent
predictions that conductance of a pure adiabatic quantum wire should
be quantized in units of $2e^2/h$ independently of interaction strength
in 1DEG, Yacoby {\em et al.} observed conspicuous deviations from the
predicted universality. Their experimental results can be summarized
as follows:

(1) At low temperature as the Fermi energy in the wire is varied 
the dc conductance shows a number of plateaus separated by the steps of
approximately
the same value 
\begin{equation} \label{sg}
G=2 g \frac{e^2}{h}, 
\end{equation}
 where $g<1$ is a temperature dependent factor
which varies from sample to sample and which may be as low as $0.75$.

(2) As temperature is increased (from $0.3 K$ up to $25 K$)
higher plateaus disappear while the factor $g$ for lower plateaus 
monotonically increases and approaches $1$.

(3) The differential conductance of the wire monotonically increases
with applied bias and for sufficiently strong biases (about 7 mV) exceeds
the universal value $2e^2/h$.

(4) Independence of the conductance on the plateau of Fermi energy,
rigid rise of plateaus when temperature is varied as well as the 
measurements
of the conductance dependence on the length of the wire suggest that
there is no disorder induced backscattering in the wire which might be
responsible for the reduced value of conductivity.

The purpose of this paper is to suggest an 
explanation of these
experimental data. We believe that two different effects 
are encountered which should be considered separately. The first effect
is responsible for  reduced values of the dc conductance at low temperatures
and of the differential conductance at low biases. In our opinion, it is
not related to the properties of 1DEG but rather is a consequence of
anomalously enhanced backscattering of electrons in 2DEG. Only this
effect will be considered in the present paper.

The second effect manifests itself in  the differential 
conductance
on a plateau in a strongly biased quantum wire exceeding the universal 
value $2e^2/h$. 
We do not consider this effect in the paper.

In the absence of impurity backscattering the conductance of a
 noninteracting 1DEG is given by 
the Landauer-B\"{u}ttiker formula \cite{LB}
\begin{equation} \label{uc}
G= 2\frac{e^2}{h}.
\end{equation}
The factor of 2 accounts for two spin orientations of electrons.
It is believed that at low temperatures and low biases 
the same universal result (\ref{uc}) holds true
in the interacting 1DEG \cite{uc}.
In the derivation of the universal conductance formula
it is assumed that electrons leaving the
wire be irreversibly absorbed by the leads. In our opinion, this condition
is not satisfied in experiments performed by Yacoby {\em et al.} 

The quantum wire of \cite{Yacoby} is formed by edge states at the boundary
of the 2DEG. At some segment of the boundary the 2DEG is depleted 
by applying the gate voltage and thereby is decoupled from the edge states.
It is this segment of the boundary that models a quantum wire.
However, the edge states extend along the whole length of the boundary.
In an ideal system the edge states would be orthogonal to the 2DEG states 
in the bulk.  Thus, if we had applied a bias voltage to the 2DEG leads
of such a system, there
would have been no current at all. Of course, in the experimental
device the 1DEG formed by edge states in the leads (outside the true
quantum wire region)  is strongly coupled to the 2DEG in the bulk.
This coupling is modeled by inserting randomly distributed scattering
centers at the boundary. 

The model suggested in \cite{Yacoby}
introduces a possibility of backscattering in the edge channels
and, presumably, scattering from one edge channel to another.
The authors of \cite{Yacoby} introduce two coefficient $\Gamma_{2D} $ and
 $\Gamma_{BS}$ characterizing
the rate of scattering to the $2DEG$ and of backscattering to the edge states.
Then by solving the Boltzman equation they show that the conductance
is reduced 

\begin{equation}
G=\frac{2e^2}{h} \ \left( 1+ \frac{2\Gamma_{BS}}{\Gamma_{2D}} \right)^{-1/2}.
\end{equation}
In our model we assume that the  backscattering to the edge states can
be neglected  $\Gamma_{BS}\approx 0$.
The reason is as follows. 
The region of the true quantum wire where the 1DEG is decoupled from the 2DEG
is quite small. It was shown experimentally that the quantum wire is
shorter than the mean free path of an electron. So, we can neglect the
impurity scattering in this region. In the region where the edge states
are strongly coupled to the 2DEG the electron-electron interaction
is irrelevant \cite{uc}. If there had been some substantial backscattering
into the edge states in this region the conductance would have been
reduced by a temperature independent factor. The experiment \cite{Yacoby}
shows a strong temperature dependence of the conductance. Therefore,
we conclude that the backscattering into the edge states is not
responsible for the conductance reduction.

Now choose a particular mode $\alpha$ in edge states. 
It is coupled to two 2DEG leads.
As we neglect electron-electron interaction in 1DEG, we can consider
the wave function of a particular electron leaving the quantum wire.
It propagates along the edge and 
is scattered to the 2DEG by the scattering centers. The $i$th scattering center
produces a contribution to the electrons wave function in
the 2DEG region 
\begin{equation}
\psi=A^{\alpha}_i \psi^{\rm in}_p(\rho, \theta).
\end{equation}
Here we denote by $\rho=\sqrt{x^2+y^2}$  the distance from 
the scattering center, the $x$-axis is orthogonal and the $y$-axis is
parallel to  the edge of the sample, the origin of the coordinate
system is placed at the scattering center, $\theta$ is the angle between
the direction of the radius vector ${\bf \rho}$ and the $x$-axis. 
The wave function $\psi^{\rm in}_p(\rho, \theta)$ is normalized so
as to correspond to the unit incident flux
and can be approximated by its
asymptotic expansion
\begin{equation} \label{psi}
\psi^{\rm in}_p(\rho, \theta)=
 \frac{\sqrt{2}}{\pi} \ \frac{e^{ip \rho}}{\rho^{1/2}}\ \cos\theta.
\end{equation}
Later we will argue why only
this long distance asymptotics is physically relevant. We have chosen the
 simplest form of
the wave function (\ref{psi}) proportional to $\cos \theta$ as we assume that
the backscattering is reasonably isotropic and this mode will give the leading
contribution into the scattering probability.

The probability
for an electron to be scattered to the 2DEG by the $i$th scattering center
is equal to $P^{\alpha}_i= |A^{\alpha}_i|^2$.
As the mean free path in the wire is quite large (experimentally it is of the
order of $10^2 k_F^{-1}$) we conclude that in average the scattering centers are 
situated
far away from each other and one  can neglect interference between 
scattering waves created by different centers. Then the total probability
for an electron to leave the edge state and to be scattered to the 2DEG at
one of the impurities is just the sum of probabilities $P_i$:
\begin{equation} \label{P}
P^{\alpha}= \sum_{i=1}^N P^{\alpha}_i = \sum_{i=1}^N |A^{\alpha}_i|^2.
\end{equation}
We assume that all modes of edge states are {\em strongly coupled} to the 2DEG.
This means that sooner or later an electron traveling along the edge is 
scattered
into the 2DEG.  Then for each edge mode the total probability $P^{\alpha} 
\approx 1$. 
We shall see that this implies the  universal behavior of plateaus when 
temperature
and bias voltage are varied.

We believe
that at low temperatures and biases there is strong backscattering of
electrons entering the 2DEG from the edge states which leads to a non-zero
probability that an electron coherently re-enters the wire after having
left it. 

Assume that the mechanism of backscattering does not mix
different edge channels.
Then the correction to the contribution into conductance of the edge mode
$\alpha$  is given by multiplying the universal factor of $2e^2/h$ by
the transmission coefficient \cite{LB}:
\begin{equation}
G=2\frac{e^2}{h} t^{\alpha}.
\end{equation}
In fact, the scattering at each center mixes different channels.
Hence, the matrix describing return amplitude for an electron leaving
some channel in the wire is not diagonal. In Born approximation the
matrix of return amplitudes is anti-Hermitian and, therefore, can be
diagonalized by a unitary transformation. 
After diagonalization
the problem for $N$-channel edge looks as 
the problem of $N$ independent channels.

In our model one can simplify the computation of transmission
coefficients $t^{\alpha}$. 
Let's introduce 
the  return probability $r^{\alpha}=1-t^{\alpha}$. As we neglect interference
between the incoming waves created by different scattering centers
one can compute $r^{\alpha}$ as
\begin{equation}
r^{\alpha} = \sum_{i=1}^N r |A^{\alpha}_i|^2,
\end{equation}
where $r$ is the return probability for an electron injected into the 2DEG at
some point at the boundary. It is reasonable to assume that $r$ does
not depend on the transversal structure of the wave function in the quantum well
and, hence, is independent of $\alpha$. By using equation (\ref{P}) one easily
sees that $r^{\alpha}$ is also $\alpha$-independent. 
At low biases the value of the return probability coefficient $r$ is completely
determined by scattering of electrons in 2DEG at the Fermi surface of 2DEG.
Hence, it should depend neither on the position of the Fermi level in 1DEG
nor on the number of filled channels. This explains independence 
of the correction factor $g$ of the number of filled channels as well
as of their filling.

Let us mention that the value of $r$ can strongly vary when the bias is 
increased
as electrons enter 2DEG at higher energies. 
A similar phenomenon is observed in measurements of tunneling conductivity, 
see e.g. \cite{Ash} and  is generally known as zero-bias anomaly.
At present a number of different mechanisms responsible for zero-bias
anomalies is known. In our opinion, the relevant mechanism is the anomalous
enhancement of the backscattering of electrons entering 2DEG with a sharp
edge due to the formation of Friedel oscillations of the electron density
near the edge of 2DEG \cite{GL}.

The variation of the density of 2DEG near its boundary is approximately
given by summing contributions to the electron density by the single-electron
states weighed with the Fermi-Dirac distribution function 
\begin{equation}
n(x)= 4 \int \frac{dk_x}{2\pi} \frac{ dk_y}{2\pi} \left( \sin^2(k_x x)-
\frac{1}{2}\right)
f(\e_{\bf k}-\mu,T).
\label{VAR}
\end{equation}
Here $ \e_{\bf k} $ is the kinetic energy of an electron with 
the wave vector
${\bf k}$, $f(\e-\mu,T)$ is the Fermi-Dirac distribution at the chemical
potential $\mu$ and the temperature $T$. 

At low temperatures the density variation (\ref{VAR}) is a slowly decaying
oscillating  function of $x$ with the oscillation period $ \pi/k_F$ where 
$k_F$ is the Fermi wave vector \cite{GL}:
\begin{equation} \label{n}
n(x) \approx \frac{\sqrt{2} k_F^2}{\pi^{3/2} (2k_F x)^{3/2}} \sin 
\left( 2k_Fx+ \frac{3\pi}{4} \right) .
\end{equation}
This is an asymptotic formula valid for  large $k_F x$. 
Using formula (\ref{n}) one can  approximate the
effective RPA scattering potential by the following local
form 
\begin{equation} \label{POT}
v(x) = (U(0) -U(2k_F)) n(x),
\end{equation}  
where $U(q)$ is the Fourier transform of the RPA screened interaction
potential. The term $U(0)$ in (\ref{POT}) is responsible for the exchange
interaction whereas $U(2k_f)$ stands for the screened Coulomb potential.
The oscillating character of the scattering potential (\ref{POT}) leads
to anomalously enhanced backscattering of the electrons entering the 2DEG
at  energies close to the Fermi energy \cite{GL}. 

As the temperature is increased the oscillating tail  of 
the density distribution becomes suppressed at the distances higher
than
$$l_T \sim \frac{1}{k_F} \sqrt{\frac{\e_F}{T}},$$
where $\e_F$ is the Fermi energy. Consequently, the anomaly in
the backscattering amplitude diminishes at higher temperatures.

The RPA expression for $U(q)$ is
\begin{equation} \label{Uq}
U(q)=U_0(q) \frac{1}{1+ \Pi(q) U_0(q)},
\end{equation}
where 
\begin{equation} \label{U0}
U_0(q)= \frac{2\pi e^2}{\epsilon |q|}
\end{equation}
is the 2-dimensional Fourier transform of the Coulomb potential,
$\epsilon \approx 10$ is the dielectric constant of AlGaAs.
The RPA polarization operator $\Pi(q)$ of 2DEG at $q=0$ is equal to
\begin{equation} \label{Pi0}
\Pi(0)= \frac{m^*}{2 \pi \hbar^2},
\end{equation}
where $m^*$ is the effective mass of an electron \cite{HLR}.
$\Pi(q)$ continuously decreases with growing $|q|$.
A simple estimate shows that in the experiment of Yacoby {\em et al.}
$\Pi(2k_F) U_0(2k_F) \leq \Pi(0) 
U_0(2k_F)=e^2k_F/(4\epsilon \e_F) \approx 0.3$.
Hence, the exchange contribution is dominating.

Now let us estimate the order of magnitude of the probability that an 
electron coherently re-enters the wire due to backscattering on Friedel
oscillations at zero temperature.  In the Born approximation the return
 probability is given by
\begin{equation} \label{R}
1-t=r=\frac{1}{(\hbar v)^2} \ |\langle {\rm out} |V|{\rm in}\rangle |^2,
\end{equation}
where the wave function $\psi_{\rm out}$ is obtained by time reversal
(complex conjugation) from the wave function $\psi_{in}$ of an electron
entering the 2DEG from the wire, $v$ is the velocity of the incoming 
electron.
As we expect that the effect comes from
the distances of the order of several $1/k_F$, we can replace the exact
wave function $\psi_{\rm in}$  by its asymptotic form (\ref{psi}).

As the effect is expected to be maximal
for electrons at Fermi energy of 2DEG, we compute the probability for
$p=k_F$.
Taking into account formulas (\ref{psi}), (\ref{n}), (\ref{POT}) and
(\ref{R}) we
obtain the following expression for the probability $r$:
\begin{equation}
\label{r}
r_0 = \kappa \left( \frac{n_0}{\pi \e_F} (U(0) - U(2k_F)) \right)^2,
\end{equation}
where $\kappa$ is a coefficient close to $1$. Let us remark that
instead of the linear dependence of the reflection coefficient
on the interaction potential derived in
the work \cite{GL} formula (\ref{r}) predicts quadratic dependence.
The difference is due to the fact that in our model the transmission
coefficient is equal to 1 to the zeroth order in scattering potential 
whereas in the context of \cite{GL} the transmission is weak.

Substituting expressions (\ref{Uq}), (\ref{U0}) and (\ref{Pi0})
into formula (\ref{r})
one gets
\begin{equation}
r_0= \frac{4\kappa}{\pi^2} \left( 1- 
\frac{ \Pi(0)U_0(2k_F)}{1+ \Pi(2k_F)U_0(2k_F)} \right)^2 .
\end{equation}
Let us remark that in the regime when $\Pi(0)U_0(2k_F)$ is small
the value of $r_0$ is not very sensitive to parameters of
the problem. If one neglects $\Pi(0)U_0(2k_F)$ one gets a universal
result $r_0 = 4\kappa/\pi^2 \approx 0.4$ . Note, however, that
for so high a return probability the Born approximation would not
be reliable anymore.

For the values of the
electron density $n_0=(1\div 2) \times 10^{11} {\rm cm}^{-2}$
and of the Fermi energy $\e_F= 10 {\rm meV}$ of the 2DEG in experiments
\cite{Yacoby} one gets $r_0 \approx 0.2$. 
This is in a good agreement with experimental values of 
$r_{\rm exp}=(10 \div 25)\times 10^{-2}$. 
Let us remark that our estimate is based on the assumption that 
the screening of the potential is purely 2-dimensional. 
This assumption is true if the wave length of the Friedel
oscillations $\pi/k_F$ is much greater than the width of the
quantum well. In experiments of Yacoby {\em et al.} 
$\pi/k_F \approx (25\div 40) {\rm nm}$ whereas the
the width of the well varies from $14 {\rm nm}$ to $40 {\rm nm}$. The largest
deviation in conductance is observed for thin wells 
$r_{\rm exp}=(20 \div 25)\times 10^{-2}$ and it decreases
when the well width grows to $r_{\rm exp}\approx 10 \times 10^{-2}$.
We think that this may happen due to cross-over from purely
2-dimensional to 3-dimensional behavior.

Analysis of the dependence of the reflection coefficient $r$ 
on the bias
shows that for strong biases $V\sim \e_F$ the asymptotic of $r$ is
given by
\begin{equation}
r(V)\sim (\e_F+V)^{-3/2}.
\end{equation}
Numerical analysis shows that this asymptotic expression works well
for $V> 0.3 \e_F$. Below this value the reflection coefficient
grows sharply as the bias approaches zero and reaches the 
finite value $r_0$ at $V=0$.

There are two different factors which make the conductivity
temperature dependent. First, the reflection coefficient 
decreases when temperature is increased because the Friedel 
oscillations are suppressed at high temperatures. This dependence
can be approximated by exponential:
\begin{equation} \label{rT}
r(T) \sim e^{-\alpha T/\e_F},
\end{equation}
where $\alpha$ is a coefficient numerically found to be
of order $1$. We expect $\alpha$ to be sensitive
to the electron density distribution near the edge. 
Second, at finite temperatures the conductivity is related
to the transmission coefficient $t=1-r(V,T)$ via
\begin{equation} \label{sT}
G= 2\frac{e^2}{h} \ 
\int d\e  \ t(\e-\mu,T) \ \frac{\partial }{\partial \mu} f(\e-\mu, T).
\end{equation}
As temperature in experiments \cite{Yacoby} is always very small in
comparison to the Fermi energy ($T\sim (1\div 20) {\rm K},
 \e_F\sim 10 {\rm mV}$)
one can approximate the dependence of $t$ on the bias $V$ by several
first terms of the Taylor expansion. As $r(V,T)$ has a maximum at $V=0$
and behaves smoothly at finite $T$, the linear term in the Taylor
expansion vanishes.
The higher
order terms result in the contributions to $G(T)$
proportional to the powers of $T$ higher than 1 and
can therefore  be neglected in view of the exponential 
suppression of $r(T)$ (\ref{rT}):
\begin{equation}
G(T)\approx 2\frac{e^2}{h} \left( 1-r_0e^{-\alpha T/\e_F}\right).
\end{equation}
This dependence of $G(T)$ is qualitatively similar to the
one obtained in experiments \cite{Yacoby}.

\vskip 0.2cm

The authors are grateful to J.Fr\"{o}hlich, D.Maslov,
T.M.Rice, A.Rodina  and A.Yacoby for useful discussions and comments. One of
the authors ( V.C.) thanks ETH, Z\"{u}rich for hospitality.


\widetext
\end{document}